\documentclass[12pt,preprint]{aastex}
\usepackage{amsmath,longtable,natbib}
\usepackage{lineno}
\usepackage{hyperref}
\shorttitle{Black-widow PSR J1544+4937}
\shortauthors{Bhattacharyya et al.}
\newcommand{\fermij}{\textsc{Fermi J1544.2+4941}}

\begin{document}

\title{GMRT discovery of PSR J1544+4937, an eclipsing black-widow pulsar identified with a \textit{Fermi} LAT source}

\author{
B.~Bhattacharyya\altaffilmark{1},
J.~Roy\altaffilmark{1},
P.~S.~Ray\altaffilmark{2},
Y.~Gupta\altaffilmark{1},
D.~Bhattacharya\altaffilmark{3},
R.~W.~Romani\altaffilmark{4},
S.~M.~Ransom\altaffilmark{5},
E.~C.~Ferrara\altaffilmark{6},
M.~T.~Wolff\altaffilmark{2},
F.~Camilo\altaffilmark{7,8},
I.~Cognard\altaffilmark{9},
A.~K.~Harding\altaffilmark{6},
P.~R.~den~Hartog\altaffilmark{4},
S.~Johnston\altaffilmark{10},
M.~Keith\altaffilmark{10},
M.~Kerr\altaffilmark{4},
P.~F.~Michelson\altaffilmark{4},
P.~M.~Saz~Parkinson\altaffilmark{11},
D.~L.~Wood\altaffilmark{12},
K.~S.~Wood\altaffilmark{2}
}
\altaffiltext{1}{National Centre for Radio Astrophysics, Tata Institute of Fundamental Research, Pune 411 007, India}
\altaffiltext{2}{Space Science Division, Naval Research Laboratory, Washington, DC 20375-5352, USA}
\altaffiltext{3}{Inter-University Centre for Astronomy and Astrophysics, Pune 411 007, India}
\altaffiltext{4}{W. W. Hansen Experimental Physics Laboratory, Kavli Institute for Particle Astrophysics and Cosmology, Department of Physics and SLAC National Accelerator Laboratory, Stanford University, Stanford, CA 94305, USA}
\altaffiltext{5}{National Radio Astronomy Observatory (NRAO), Charlottesville, VA 22903, USA}
\altaffiltext{6}{NASA Goddard Space Flight Center, Greenbelt, MD 20771, USA}
\altaffiltext{7}{Columbia Astrophysics Laboratory, Columbia University, New York, NY 10027, USA}
\altaffiltext{8}{Arecibo Observatory, Arecibo, Puerto Rico 00612, USA}
\altaffiltext{9}{ Laboratoire de Physique et Chimie de l'Environnement, LPCE UMR 6115 CNRS, F-45071 Orl\'eans Cedex 02, and Station de radioastronomie de Nan\c{c}ay, Observatoire de Paris, CNRS/INSU, F-18330 Nan\c{c}ay, France}
\altaffiltext{10}{CSIRO Astronomy and Space Science, Australia Telescope National Facility, Epping NSW 1710, Australia}
\altaffiltext{11}{Santa Cruz Institute for Particle Physics, Department of Physics and Department of Astronomy and Astrophysics, University of California at Santa Cruz, Santa Cruz, CA 95064, USA}
\altaffiltext{12}{Praxis Inc., Alexandria, VA 22303, resident at Naval Research Laboratory, Washington, DC 20375, USA}

\affil{}

\begin{abstract} 
Using the Giant Metrewave Radio Telescope (GMRT) we performed deep observations to search for radio 
pulsations in the directions of unidentified \textit{Fermi} Large Area Telescope (LAT) $\gamma$-ray sources. We report the 
discovery of an eclipsing black-widow millisecond pulsar, PSR J1544+4937, identified with the 
un-cataloged $\gamma$-ray source \fermij. This 2.16 ms pulsar is in a 2.9 hours compact circular orbit with a very low-mass  
companion ($M_c > 0.017 M_{\odot}$). At 322 MHz this pulsar is found to be eclipsing for 13\% of its 
orbit, whereas at 607 MHz the pulsar is detected throughout the low-frequency eclipse phase. Variations in the eclipse ingress phase are observed, indicating a clumpy and variable eclipsing medium.  Moreover, additional short-duration absorption events are observed around the eclipse 
boundaries. Using the radio timing ephemeris we were able to detect $\gamma$-ray pulsations from this pulsar, confirming it as the source powering the $\gamma$-ray emission.
\end{abstract}

\vskip 0.6 cm

\section{Introduction}
\label{sec:intro}
The Large Area Telescope \citep[LAT;][]{atwood09} aboard the \textit{Fermi Gamma-ray Space Telescope}
has discovered a large number of $\gamma$-ray point sources, of which many are unidentified or even 
unassociated with any known potential counterpart \citep{Ackermann2012}. The LAT can localize most of these sources well enough that they can be covered in a single pointing with the large primary beams of radio telescopes at low-frequencies, allowing them to be searched efficiently. Targeted radio searches of unassociated LAT point sources by the \textit{Fermi} Pulsar Search Consortium (PSC) have resulted in the discovery 
of 43 radio millisecond pulsars \citep[MSPs;][]{Ray12}. 
MSPs are thought to evolve from normal pulsars in binary systems via transfer of 
angular momentum from companions. Thus, the majority of MSPs are naturally expected to be 
in binaries ($\sim$ 83\% being the binary fraction for MSPs in the Galactic field\footnote{\url{http://astro.phys.wvu.edu/GalacticMSPs/}}). Binary systems where the pulsar wind 
evaporates the companion are one way to form isolated MSPs. 
Such systems where the interaction is ongoing are called black-widow (BW) pulsars. Many exhibit long eclipses ($\sim$ 10\% 
of the orbital period, apparently larger than the companion's Roche lobe) that are believed to be caused by the material blown from the 
very low mass companion ($M_c \ll 0.1 M_{\odot}$) by the pulsar wind.
There were two such eclipsing BW systems in the Galactic field known before 
the launch of \textit{Fermi} $-$ PSR B1957$+$20 \citep{Fruchter88} and PSR J2051$-$0827 
\citep{Stappers96}. The BW pulsars are found to have higher values of spin-down energy-loss rate
($\dot{E} \sim$ 10$^{34}$ erg s$^{-1}$) compared to other MSPs, making these systems good candidates for pulsed 
$\gamma$-ray emission \citep{Mallory2011}. 
Among 43 new MSPs found in \textit{Fermi}-directed searches there are at least 10 BWs \citep{Ray12}.
This Letter describes the discovery and follow-up study of an eclipsing BW MSP, 
J1544+4937, with the GMRT.

\section{Observations and search analysis}  
\label{sec:obs_analysis}

As a part of the PSC search effort, we observed mid- and high-Galactic-latitude unassociated 
\textit{Fermi} point sources 
with the GMRT at 607 MHz. The GMRT Software Back-end \citep[GSB;][]{Roy10} produces 
simultaneous incoherent and coherent filter-bank outputs of 512$\times$0.0651 MHz sampled every 61.44 $\mu$s. 
The wider incoherent beam of the GMRT (40\arcmin~at 607 MHz) can easily cover error-circles associated 
with the \textit{Fermi} sources. In addition a coherent beam that is $3\times$ more sensitive and narrower (1.5\arcmin~at 607 MHz 
using the central core of the GMRT) can be useful if the pulsar happens to be near the pointing center.

One of the targets was \fermij, a $\gamma$-ray source from an unpublished internal source list created 
by the LAT Collaboration using 18 months of data in preparation for the 2FGL catalog \citep{Nolan2012}. The source 
location (J2000) from that analysis (and used for our telescope pointing) was R.A. = 236\degr.074, Decl. = 49\degr.695, with 
a 95\% confidence error-circle of radius 9.5\arcmin. This source is very weak, with a likelihood test statistic \citep[TS;][]{Mattox1996} of 26.2 
 in the 18 month analysis, and did not make the significance cut to be included in the 2FGL catalog itself.

We processed the data on an IUCAA HPC cluster with Fourier-based acceleration search methods using PRESTO \citep{Ransom02}. 
We investigated trial dispersion measures 
(DMs) ranging from 0 pc cm$^{-3}$ up to 350 pc cm$^{-3}$. A linear drift of up to 200 Fourier-frequency bins for the 
highest summed harmonic was allowed. The powerline, 50 Hz, and its subsequent harmonics were
excised. Using parameters of 32 MHz bandwidth, 10\% duty-cycle, incoherent array gain of 2.3 K/Jy, 
for 30 minutes of observing, we estimate the search sensitivity as (92K $+T_\mathrm{sky}$)/(335K) mJy for a 5$\sigma$ 
detection at 607 MHz. Considering $|b|>$ 5\degr, where  $T_\mathrm{sky}$ $\sim$ 10--45 K, our search sensitivity is 0.3--0.4 mJy. 

In a 30-minute pointing on 2011 February 1, towards \fermij\, we discovered a binary MSP of period 2.16 ms with
significant acceleration of 2.25 m s$^{-2}$ at a DM of 23.2 pc cm$^{-3}$. 

\section{Follow-up timing} 
\label{sec:timing}

\begin{figure}[htb] 
\includegraphics[width=4.0in]{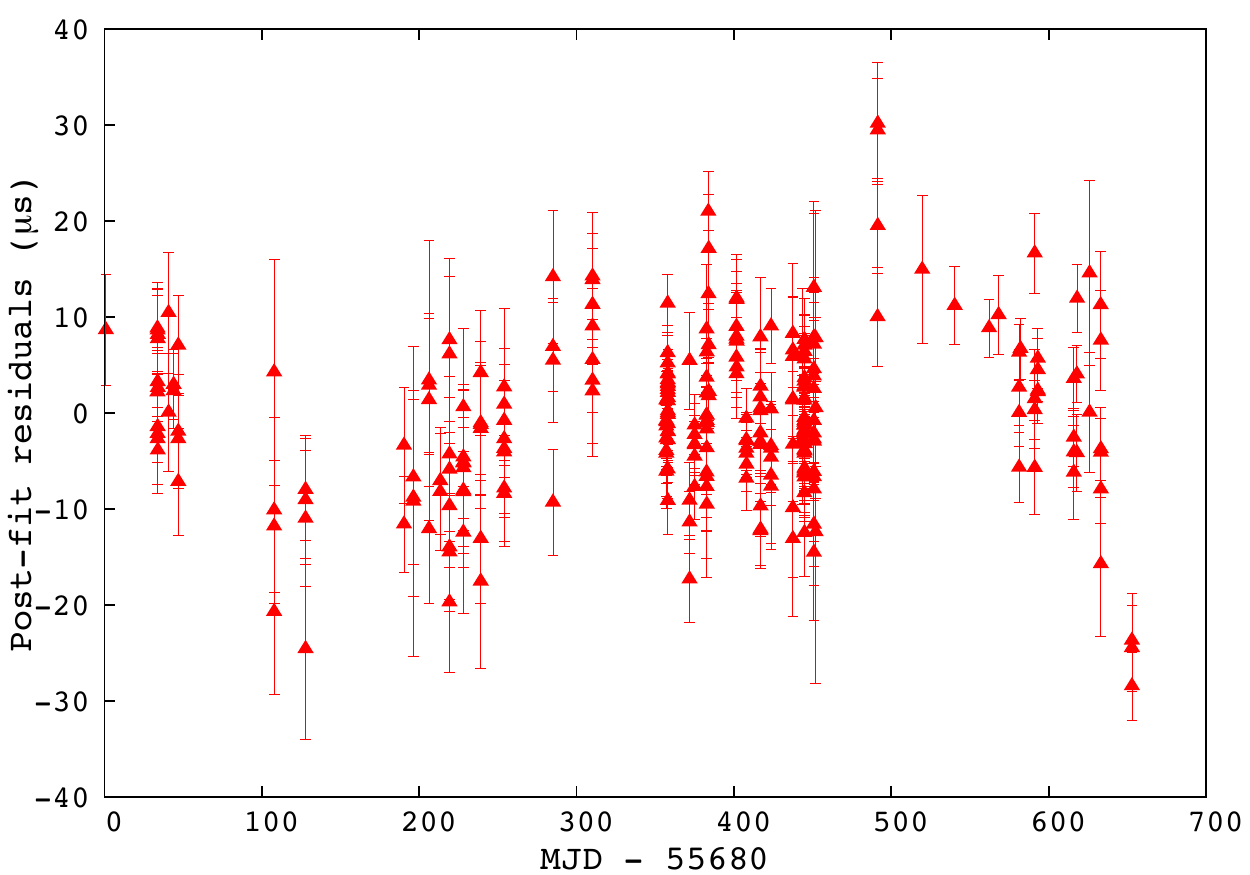}
\caption{Post-fit timing residuals of J1544$+$4937 considering non-eclipsing binary phases. \label{fig:residual}}
\end{figure}

We localized J1544$+$4937 with an accuracy of 5\arcsec~(positions listed in Table~\ref{tab:params}) using continuum 
imaging for the full GMRT array followed by multi-pixel beamforming \citep{Roy12}, which allowed us to have sensitive 
follow-up studies using the coherent array.  
We estimate a flux of 5.4 mJy at 322 MHz, and a spectral index of $-$2.3.
We started the regular timing campaign for J1544$+$4937 in April 2011 at 322 MHz with the same coherent filter-bank.
With the derived position from the multi-pixel search and an \textit{a priori} binary model predicted by \cite{Bhattacharyya08}, 
we obtained phase-connected time-of-arrivals (TOAs) from TEMPO\footnote{\url{http://tempo.sourceforge.net}}, using the
JPL DE405 solar system ephemeris \citep{Standish04}. 
The binary timing model used is ELL1 \citep{Lange01}, since J1544$+$4937 is in a very low eccentricity system. 
This MSP is in a very compact binary with an orbital period of 2.9 hours. We derive a minimum companion mass (for 90\degr\ orbital 
inclination) of 0.017 M$_{\odot}$ using the Keplerian mass function, assuming a pulsar mass
of 1.4 M$_{\odot}$.
J1544$+$4937 is eclipsed for about 13\% of the orbit at 322 MHz (Sec.\ref{sec:eclipse_ch}). The best-fit timing model 
(MJD 55680.927--56332.90) is obtained excluding the TOAs around the eclipse phase (0.05$-$0.35). We achieved a 
post-fit rms timing residual of 6.9 $\mu$s from 652 days of timing (Fig.~\ref{fig:residual}). There are still un-modeled residuals,
which can be partially absorbed by proper motion fit. However the inclusion of proper motion reduces the LAT detection significance,
indicating that more timing data are required to improve the model. 
We estimate a precise DM equal to 23.2258(11) pc cm$^{-3}$ by a timing fit using 322 and 607 MHz TOAs from non-eclipsing binary phases. Ephemeris, 
position and derived parameters are listed in Table~\ref{tab:params}. 

\begin{deluxetable}{ll}
\tabletypesize{\scriptsize}
\tablewidth{0pt}
\tablecaption{Parameters of J1544$+$4937
\label{tab:params}}
\tablehead{
\colhead{Parameter} & \colhead{Value\tablenotemark{a}}
}
\startdata
  \hline
  \multicolumn{2}{c}{Interferometric position\tablenotemark{b}} \\
  \hline
Right ascension (J2000)\dotfill & 15$^\mathrm{h}$44$^\mathrm{m}$04\fs166$\pm$0\fs3\\
Declination (J2000)\dotfill     & +49\degr37\arcmin57\farcs45$\pm$4\farcs7\\
Offset from survey beam center\dotfill  & 4.3\arcmin    \\
  \hline
  \multicolumn{2}{c}{Parameters from radio timing} \\
  \hline
Right ascension (J2000)\dotfill & 15$^\mathrm{h}$44$^\mathrm{m}$04\fs48722 (2)\\
Declination (J2000)\dotfill     & +49\degr37\arcmin55\farcs2545 (2) \\
Position Epoch (MJD)\dotfill    & 51544.0      \\
Pulsar period $P$ (ms)\dotfill  & 2.15928839043289 (5) \\
Pulsar frequency $f$ (Hz)\dotfill  & 463.11553585462 (1)\\
Frequency derivative $\dot{f}$ (Hz s$^{-1}$)\dotfill & $-$6.29 (1)$\times$10$^{-16}$ \\
Period Epoch (MJD)\dotfill  & 56007.0\\
Dispersion measure $DM$ (cm$^{-3}$ pc)\dotfill & 23.2258 (11)\\
Binary model\dotfill & ELL1\\
Orbital period $P_{b}$ (days)\dotfill       & 0.1207729895 (1) \\
Projected semi-major axis $x$ (lt-s)\dotfill & 0.0328680 (4) \\
Epoch of ascending node passage $T_{ASC}$ (MJD)\dotfill & 56124.7701121 (2)\\
Span of timing data (MJD)\dotfill & 652 \\
Number of TOAs\dotfill & 280 \\
Post-fit residual rms ($\mu$s)\dotfill & 6.9\\
Reduced chi-square\dotfill & 2.7 \\
 \hline
  \multicolumn{2}{c}{Derived parameters} \\
  \hline
Mass function $f$ (M$_{\odot}$)\dotfill  & 0.0000026132\\
Min companion Mass $m_{c}$ (M$_{\odot}$)\dotfill & 0.017\\
DM distance\tablenotemark{c} (kpc)\dotfill & 1.2\\
Flux density at 322 MHz (mJy)\dotfill & 5.4\\
Flux density at 607 MHz (mJy)\dotfill & 1.2\\
Spectral index\dotfill & $-$2.3\\
Surface magnetic field $B_{s}$ ($10^{8}$ G)\dotfill & 0.805 (1) \\
Spin down luminosity  \.{E} (10$^{34}$ erg s$^{-1}$)\dotfill & 1.150 (8) \\
Characteristic age $\tau$ (Gyr)\dotfill & 11.65 (3) \\
 \hline
  \multicolumn{2}{c}{$\gamma$-ray parameters\tablenotemark{d}} \\
  \hline
Photon flux ($>0.1$ GeV, cm$^{-2}$ s$^{-1}$)\dotfill & 1.6(8) $\times 10^{-9}$\\
Energy flux ($>0.1$ GeV, erg cm$^{-2}$ s$^{-1}$)\dotfill & 2.1(6) $\times 10^{-12}$\\
Luminosity, $L_\gamma/f_\Omega$ ($>0.1$ GeV, erg cm$^{-2}$ s$^{-1}$)\dotfill & 3.6 $\times 10^{32}$\\
Efficiency, $\eta_\gamma/f_\Omega$ ($>0.1$ GeV)\dotfill & 0.03 \\
\enddata
\tablenotetext{a}{Errors in the last digit are in parentheses.}
\tablenotetext{b}{\cite{Roy12}}
\tablenotetext{c}{\cite{Cordes02}}
\tablenotetext{d}{phase-averaged}
\end{deluxetable} 

\begin{figure}[htb]
\includegraphics[width=3.5in]{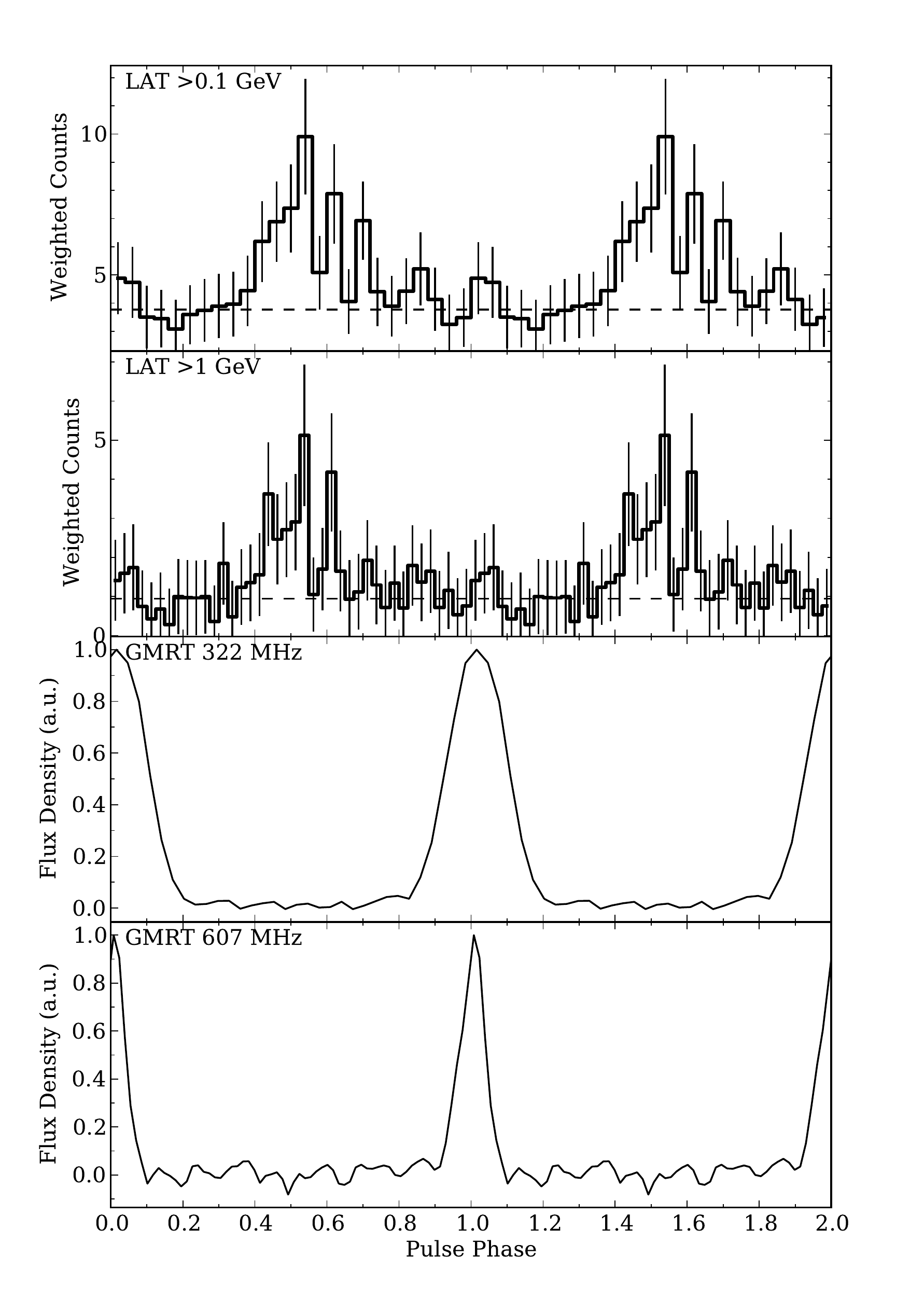}
\caption{Phase-aligned radio and LAT $\gamma$-ray light-curves of J1544+4937. The horizontal dashed lines are an estimate of the background level from sources other than the pulsar. The 322 MHz profile is broadened by incoherent dedispersion across the channel bandwidth of 0.0651 MHz, introducing $\sim 373 \mu$s smearing. \label{fig:lc}}
\end{figure}
\section{$\gamma$-ray pulsations} 
\label{sec:gammaray}

\begin{figure}[htb]
\includegraphics[width=6.0in]{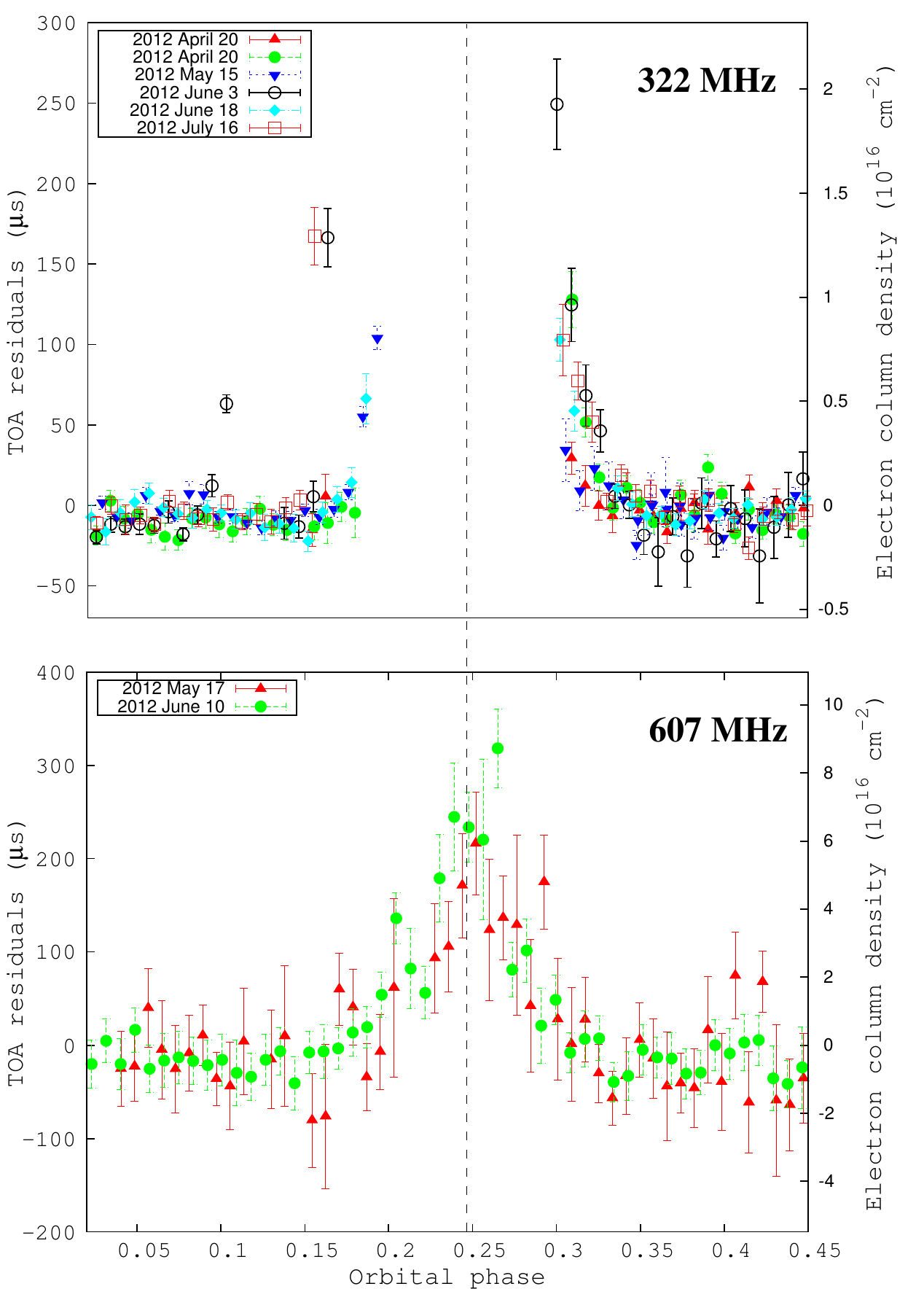}
\caption{Variation of timing residuals and electron column density (TOAs of 90 s time resolution) around eclipse phase
at 322 MHz (top) and 607 MHz (bottom). \label{fig:pulsars_all_group}}
\end{figure}

\begin{figure}[htb]
\includegraphics[width=6.0in]{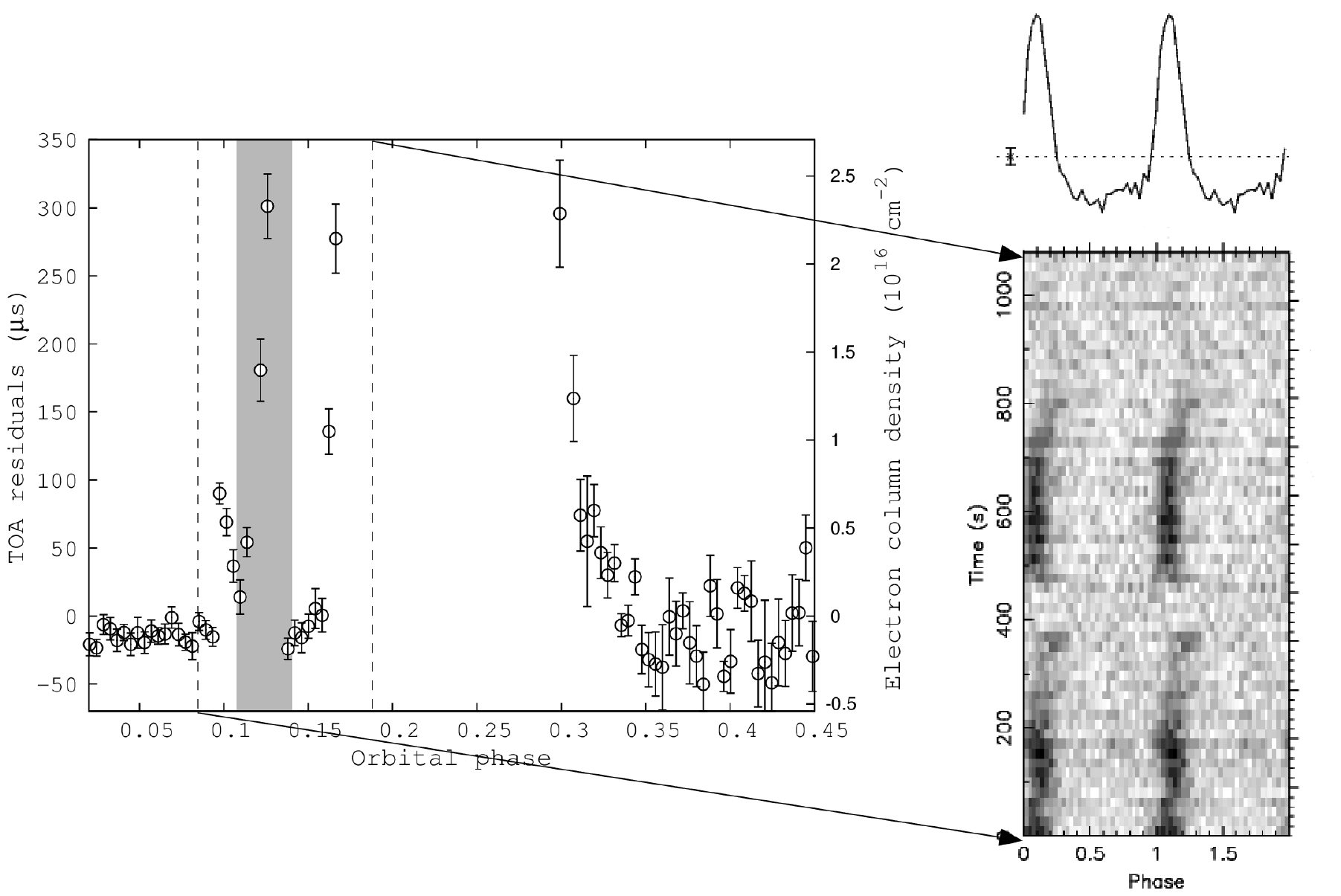}
\caption{Example of additional short-duration absorption beyond the eclipse phase at 322 MHz on 2012 June 3.
Left: variation of timing residuals and electron column density at 42 s time resolution. Right: phaseogram
showing short-duration absorption. \label{fig:short_eclipse}}
\end{figure} 


The radio timing position of J1544+4937 is 4.3\arcmin\ from the LAT localization of \fermij. 
This is well within the radius of the 95\% confidence error circle, suggesting an association 
between the pulsar and the LAT source, which we sought to confirm with detection of pulsations.  
In addition, the $\gamma$-ray detectability metric \.{E}$^{1/2}$/$d^2$ for J1544$+$4937 is 
7.4 $\times$ 10$^{16}$ erg$^{1/2}$ kpc$^{-2}$ s$^{-1/2}$, which is comparable to other $\gamma$-ray detected MSPs \citep[Fig.~12 of][]{abdo10}. 

Because of the low significance of the LAT source, a phase-averaged spectral analysis of the region 
is required to optimize the sensitivity of the H-test for pulsed significance using photon probability weighting \citep{Kerr2011}.
We performed a binned likelihood spectral analysis of the region using the \textit{Fermi} Science 
Tools\footnote{\url{http://fermi.gsfc.nasa.gov/ssc/data/analysis/software/}}. We selected LAT data with 
reconstructed energies in the range 100 MeV to 20 GeV collected between 2008 August 4 and 2013 February 6. 
We used the P7SOURCE\_V6 instrument response functions and excluded events at a zenith angle $>100$\degr\ 
and times when the LAT was in the SAA or the rocking angle was greater than 52\degr. We modeled\footnote{\url{http://fermi.gsfc.nasa.gov/ssc/data/access/lat/BackgroundModels.html}} a region of radius 15\degr\ using a source list from the 2FGL catalog with 
an additional source added at the location of the pulsar. We modeled the pulsar with an exponentially 
cutoff power-law spectrum.  Because it is so faint, we fitted only for the power-law index and normalization, 
keeping the cutoff energy fixed at 2.5 GeV, a value 
typical of other MSPs \citep{abdo10}. The spectral model included both the isotropic and Galactic diffuse 
contributions with normalizations free in the fit.  We detected the source with a TS of 38.2, corresponding to $\sim 6\sigma$. The spectral index and cutoff energy are very poorly constrained for this weak source, so 
to estimate the uncertainty on the integrated flux we repeated the spectral fit with the cutoff frozen at 0.7, 
2.5, and 5.0 GeV, and with the index frozen at $1.3$ (the average value for LAT-detected MSPs) and the cutoff free.  
We find that the uncertainty due to the spectral shape is about $3\times 10^{-10}$ cm$^{-2}$ s$^{-1}$ for the 
photon flux and $2\times 10^{-13}$ erg cm$^{-2}$ s$^{-1}$ for the energy flux.  These are smaller than the 
statistical errors for this faint source and are added in quadrature to compute the errors on the fluxes 
reported in Table \ref{tab:params}. Using the fitted energy flux ($G$), and the DM distance ($d$), we compute 
the $\gamma$-ray luminosity $L_\gamma = 4 \pi f_\Omega G d^2$ and efficiency $\eta = L_\gamma/\dot{E}$, where 
the beaming factor $f_\Omega$ is assumed to be 1 \citep{Watters09}.

Using this model for the continuum emission from the region, we search for pulsed $\gamma$-ray emission by selecting events within
2\degr\ of the source and computed the probability that they originated 
from the source with the \textit{Fermi} Science Tool \texttt{gtsrcprob}.  Using these probabilities as weights and 
pulse phases computed using the radio timing model, we computed a weighted H-test of 37.1, corresponding to a detection 
significance of 5.1 $\sigma$ and confirming that J1544+4937 is indeed a $\gamma$-ray pulsar and identifying the 
LAT source as this MSP. The LAT weighted light-curves in two energy bands, phase aligned to the radio profiles, are 
shown in Fig. \ref{fig:lc}.  For this faint source, it is difficult to determine the peak multiplicity, but 
the $\gamma$-ray emission seems to be mainly between phases 0.4 and 0.6, relative to the single radio peak. Fitting 
a single gaussian peak to the light-curve gives a peak at phase $0.54 \pm 0.02$ with a width of $0.25 \pm 0.05$.
 
\section{X-ray and optical observations}
\label{sec:xray_optical}

We analyzed 5 {\it Swift}/XRT observations (totaling 9.3 ks) of this field from 2013 March 19--24. No X-ray counterpart is 
detected, providing a flux upper limit of $\sim$ 1.7$\times$10$^{-14}$ ergs cm$^{-2}$ s$^{-1}$ for 0.3$-$10 keV.

We have checked for an optical counterpart for J1544$+$4937 using archival
SDSS data. This field was imaged \citep{Adelman08} on MJD 52046.294, at binary
phase 0.57.  No optical counterpart was detected. The strongest upper
limits were $g^\prime > 22.5$, $r^\prime >22.0$ and $i^\prime >21.5$ (90\% CL).
We also obtained a 300s H$\alpha$ image with the MiniMo camera on the 3.5m WIYN telescope
on MJD 55975.521 \citep{Brownsberger13}. No stellar counterpart or extended emission was
seen with an H$\alpha$ flux upper limit of $2.5 \times 10^{-16}$ erg\,cm$^{-2}\,s^{-1}$ (90\% CL)
for a compact $< 5^{\prime\prime}$ nebula.

\cite{Breton13} have obtained optical photometry of the heated companions of a
number of {\it Fermi} MSPs. They find a typical heating efficiency
$\eta \sim 0.15$. For systems with a very strong wind $\eta$ approaches unity
(e.g. J1810$+$1744, \cite{Breton13}; J1311$-$3430, \cite{Romani12}).
\cite{Breton13} give an estimate for the irradiation temperature
$T_{Irr}=(\eta {\dot E}_{SD}/4\pi\sigma a^2)^{1/4}$, which is 4370$\,K$ considering $\eta =0.15$
and 6360$\,K$ considering $\eta = 0.6$ for J1544$+$4937.
If we assume that the deep, variable radio eclipses imply a near Roche lobe filling companion,
and consider a typical inclination angle $=60^\circ$, an unheated (backside)
companion temperature of 2500\,K and $\eta= 0.15$, we can use the binary light-curve 
synthesis program `Icarus' \citep{Breton11} to predict magnitudes at
phase $\phi_B=0.57$ of $g^\prime \approx 25.2$, $r^\prime \approx 23.9$ and
$i^\prime \approx 23.1$ (magnitudes at maximum are 23.0, 21.9 and 21.5, respectively).
Thus the observed SDSS limits are not constraining. However if
the irradiation efficiency is higher, the fluxes can be detectable; for example
with $\eta= 0.6$ we expect $g^\prime \approx 22.4$, $r^\prime \approx 21.8$ and
$i^\prime  \approx 21.6$ at $\phi_B=0.57$. These are comparable to our observed
magnitude limits, so higher efficiencies are ruled out unless the Roche lobe filling
factor is small or the source distance is larger. These estimates also depend weakly
on the observer inclination and the secondary unheated temperature and composition

Our H$\alpha$ limit for J1544$+$4937 corresponds to $<0.15$ of the highest
surface brightness $\sim 5^{\prime\prime}$ patch of the B1957$+$20 bow-shock.
The average flux ratio is even more constraining, with an
upper limit of $\sim 0.03$ of the total bow-shock flux. However, given the
small fraction of MSPs that show optical bow-shocks (likely due to the small
filling factor of the neutral interstellar medium), the non-detection at such a 
large distance from the Galactic plane (b$\sim$ 50\degr) is not surprising.
 
\section{Eclipse characteristics}
\label{sec:eclipse_ch}
Fig.~\ref{fig:pulsars_all_group} presents the timing residuals (and electron column densities) around the eclipse phase at 
322 and 607 MHz, with 90s time resolution. The effect of the eclipses is generally 
seen from 0.18 to 0.31 orbital phase (eclipse-zone hereafter) at 322 MHz. The eclipses are 
centered at binary phase 0.24 with a duration of around 22 minutes.
We estimate the radius of the companion's Roche lobe, R$_L$ \citep{Eggleton83},\\
\begin{equation}  
         {R_{L}}=\frac{0.49 a q^{2/3}}{0.6q^{2/3}+ \ln(1+q^{1/3})} \sim 0.13 R_{\odot}
\label{eqn1}
\end{equation}
where $q=m_c/m_p$ is the mass ratio of the companion and the pulsar, and $a$ is the separation of the companion from the pulsar 
($a \sim 1.2 R_{\odot}$ for J1544$+$4937, indicating an extremely compact binary).
The opaque portion of the companion's orbit is 0.98 R$_{\odot}$, much larger than R$_L$ of 0.13~R$_{\odot}$, so the 
volume occupied by the eclipsing material is well outside the companion's Roche lobe, and thus is not gravitationally bound to the companion. 
This confirms that this binary is a BW, where the pulsar is ablating its companion, creating a significant amount of intrabinary 
material that obscures the pulsar's emission.

Our sample consisted of six eclipses at 322 MHz where the pulsar emission was fully obscured by companion and its wind 
and two 607 MHz observing sessions covering the full orbit (Fig.~\ref{fig:pulsars_all_group}). At 607 MHz we detect the 
MSP throughout the 322 MHz eclipse-zone. However, we observe a flux fading at 607 MHz near the superior conjunction 
(orbital phase 0.24).
Significant delays in pulse arrival times are observed at 607 MHz during the eclipse-zone at 322 MHz. Maximum delay 
in pulse arrival time at 607 MHz near the eclipse superior conjunction is around 300 $\mu$s, which corresponds to an increase 
in DM of 0.027 pc cm$^{-3}$ and an added electron density, N$_e$ of 8$\times$10$^{16}$ cm$^{-2}$. 

\citet{Thompson94} (T94 hereafter) elucidate a collection of eclipse mechanisms. According to them, eclipsing due to refraction of the radio beam 
demands an order of magnitude higher group delay ($\sim$ few tens of milliseconds) than we observe for J1544$+$4937 
(250 $\mu$s at 322 MHz egress). Using N$_e \sim$8$\times$10$^{16}$ cm$^{-2}$ observed during superior conjunction at 607 MHz and 
absorption length about twice the size of the eclipse-zone, according to equation 11 of T94 we find that free-free absorption is possible 
(absorption optical depth $\tau_{ff} > 1$) if the plasma temperature T$\leq 4 \times {f_{cl}}^{2/3}$ K, where ${f_{cl}}^{2/3}$
is the clumping factor. This demands either a very low temperature or a very high value of the clumping factor, both of which are not physically
achievable. Eclipsing by pulse smearing (due to the increase of N$_e$ along the line-of-sight) can be ruled out, as the excess electron column
density inferred from 607 MHz predicts 373 $\mu$s smearing of pulses at 322 MHz near superior conjunction (considering 
incoherent dedispersion), which is less than one fifth of pulse period. Since J1544$+$4937 has a narrow main-pulse, and no significant 
profile evolution is apparent at the eclipse boundary, pulse broadening due to scattering can reduce the detectability but cannot explain 
the eclipse. In addition, since J1544$+$4937 is relatively weak, nearby and has a shallower spectrum than B1957$+$20, the 
expected induced Compton scattering optical depth is much less than one (equation 26 of T94). Another eclipse mechanism considered by 
T94 is cyclotron-synchrotron absorption of the radio waves by non-relativistic/relativistic electrons, which requires a magnetic field 
in the vicinity of the companion. We calculate a magnetic field B $\sim$ 11 G and corresponding cyclotron absorption 
frequency $\sim$ 31 MHz (equations 35, 37 of T94, assuming a moment-of-inertia of 10$^{45}$ g cm$^2$). Thus 322 and 607 MHz will 
correspond to 10th and 20th harmonics of the cyclotron resonance. For a fixed temperature, the optical depth for cyclotron absorption drops with 
harmonics, which may explain the lack of absorption seen at 607 MHz. 
A modified model is proposed by \cite{Khechinashvili00} based on kinematic treatment of cyclotron damping, assuming white-dwarf companions 
with reasonably strong surface magnetic fields. Further observations over a wider radio spectrum may help to 
investigate the frequency-dependent degree of damping predicted by this model. 

We observe a temporal variation of the ingress phase at 322 MHz. For two eclipses, the ingress phase is shifted to 0.16 (from 0.18 
for the other four eclipses), whereas there are no apparent shifts in the egress phase. This corresponds to an increase of the opaque portion by 
0.15 R$_{\odot}$ and N$_e >$1.2$\times$10$^{16}$ cm$^{-2}$ at the ingress boundary. Such asymmetric increase of eclipse duration 
may indicate that our line-of-sight is probing a wind zone where there is systematic outflow of eclipse material.

We also observe strong phase modulations and additional short-duration absorptions at ingress and egress, in time-series 
data at higher resolution. The durations of these features are in general around 10--20 s, and hence they are not seen in 
Fig.~\ref{fig:pulsars_all_group}. However, in one of the observing epochs these modulations lasted longer $-$ phase modulation of 
duration 100 s, followed by a short-duration absorption of $\sim$ 180 s, then regular emission resumes for 500 s, after which the 
eclipse starts (Fig.~\ref{fig:short_eclipse}). Fragmented blobs of plasma randomly oriented around eclipsing zone, obscuring 
radiation from the pulsar, can explain these short-duration absorptions.

\section{Discussion} 
\label{sec:discussion}

We report the GMRT discovery of an eclipsing BW pulsar, at the position of an unassociated LAT source,
\fermij. This is the first Galactic field MSP discovered at the GMRT. The detection of pulsed $\gamma$-rays from this pulsar 
demonstrates it as the source powering \fermij. Due to the limited significance of the source in $\gamma$-rays additional data are needed before conclusions on peak multiplicity and system geometry can be drawn. 
The implied efficiency ($\sim$ 3\%) of converting spin-down energy into $\gamma$-rays is typical of LAT-detected MSPs.
 This is the first discovery of a radio MSP in a LAT source fainter than the 2FGL catalog limit \citep{Ray12}. Since the 
radio pulsar is relatively bright, this provides strong justification to continue these searches as new LAT sources 
are revealed in analyses of longer datasets.  The radio flux is uncorrelated with the $\gamma$-ray flux \citep{Ackermann2012}, 
so even faint new LAT sources can harbor bright radio MSPs.

Eclipsing BW pulsars have the potential of providing information on the evolutionary connection between the low-mass X-ray 
binaries and isolated MSPs. 
With long term monitoring of this pulsar we aim to estimate $\dot P_{b}$ and its higher derivatives, which can provide an estimate 
of the life span of the system.
\citet{Bates11} noted that for BW systems the measured value of $\dot {E}$/a$^2$
is an order of magnitude higher than for other MSP binaries, indicative of greater energy flux needed to ablate the companion. For 
J1544$+$4937 we calculate $\dot {E}$/a$^2$ $\sim$ 1.5 $\times$10$^{33}$ erg lt-s$^{-2}$s$^{-1}$,  
which is similar to other BW systems in the Galactic field. 
Dual frequency observations presented in this paper suggest that cyclotron 
absorption by the plasma formed via interaction of the pulsar wind with ablated material can obscure the pulsed emission. 
However, exploring the radio spectrum on either side to probe the reduced/increased opaqueness of the stellar wind during the 
eclipse phase may provide better insight into the plausible eclipse mechanism. 

\acknowledgments

The \textit{Fermi} LAT Collaboration acknowledges support from a number of agencies and institutes for both development and the operation of the LAT as well as scientific data analysis. These include NASA and DOE in the United States, CEA/Irfu and IN2P3/CNRS in France, ASI and INFN in Italy, MEXT, KEK, and JAXA in Japan, and the K.~A.~Wallenberg Foundation, the Swedish Research Council and the National Space Board in Sweden. Additional support from INAF in Italy and CNES in France for science analysis during the operations phase is also gratefully acknowledged. We acknowledge support of telescope operators of the GMRT, which is run by the National Centre for Radio Astrophysics of the Tata Institute of Fundamental Research.
We thank the {\it Swift} team at Pennsylvania State University, especially Abe Falcone. We acknowledge help of C. Cheung in interpreting the XRT data.We thank D. Thompson and T. Johnson for their comments and R. Breton for a discussion of heating fluxes.


\end{document}